\documentclass{emulateapj}
\usepackage{apjfonts}

\slugcomment{Accepted for publication in ApJ Letter}

\shorttitle{Synchrotron X-ray Variability in Cas A}
\shortauthors{Uchiyama et al.}

\begin{document}


\title{Fast Variability of  Nonthermal X-Ray Emission
in Cassiopeia~A: \\ Probing Electron Acceleration in Reverse-Shocked Ejecta}


\author{Yasunobu Uchiyama, \altaffilmark{1} \& Felix A. Aharonian \altaffilmark{2,3}}
\altaffiltext{1}{Department of High Energy Astrophysics, ISAS/JAXA, 3-1-1 Yoshinodai, Sagamihara, Kanagawa, 229-8510, Japan;
uchiyama@astro.isas.jaxa.jp}
\altaffiltext{2}{Dublin Institute for Advanced Studies, 5 Merrion Square, Dublin 2, Ireland}
\altaffiltext{3}{Max-Planck Institut f\"ur Kernphysik, 
PO Box 103980, D69029 Heidelberg, Germany}

\begin{abstract}
Recent discovery of the year-scale variability in the synchrotron 
X-ray emission of the supernova remnant (SNR) RX J1713.7$-$3946 
has initiated our study of  multi-epoch X-ray images and spectra
of the young SNR Cassiopeia A based on the \emph{Chandra}
archive data taken in 2000, 2002, and 2004. 
We have found year-scale time variations in the X-ray intensity 
for a number of X-ray filaments or knots 
associated with the reverse-shocked regions. 
The X-ray spectra of the variable filaments are characterized by 
a featureless continuum, 
and described by a power law with a photon index within 1.9--2.3. 
The upper limits on the iron K-line equivalent width 
are 110 eV, which favors a synchrotron origin of the X-ray emission. 
The characteristic variability timescale of 4 yr can 
be explained by 
the effects of fast synchrotron cooling and diffusive 
shock acceleration  with 
a plausible magnetic field of 1 mG.
The X-ray variability provides a new effective way of studying 
particle acceleration at supernova shocks. 
\end{abstract}


\section{Introduction}

The young ($\sim 330\ \rm yr$ old) 
supernova remnant (SNR) Cassiopeia~A (Cas~A)
is a unique astrophysical laboratory for studying high-energy phenomena 
in SNRs because of its brightness across 
the entire electromagnetic spectrum from radio \citep[e.g.,][]{Baars77} 
to very-high-energy gamma-rays \citep{Aha01,Magic}.
While the X-ray emission of Cas~A below 3 keV is dominated by line emissions 
from the thin thermal plasma of the shocked ejecta, 
as demonstrated by observations with \emph{ASCA} \citep{Holt94}, 
a prominent nonthermal  tail  
is known to be present and dominant above 10 keV \citep{Allen97,Fava97}.

Nonthermal  X-rays appear to be emitted mainly from fragmented filaments 
or knots in reverse-shocked regions. 
The bulk of the hard X-ray emission in 10--12 keV 
 from Cas~A comes from 
such components; the total emission in the outer peripheral filaments 
found by \emph{Chandra} \citep{Hugh00,VL03} 
is only a small fraction of the total 4--6 keV continuum flux  \citep{Blee01}.  
The interpretation of the nonthermal component in reverse-shocked ejecta remains 
an open issue \citep[e.g.,][]{Lam01}.

In this Letter, we present the analysis of multi-epoch 
\emph{Chandra} imaging and spectrometric data of Cas~A. 
This work was motivated by our recent discovery of time variability in 
the synchrotron X-ray emission of SNR RX~J1713.7$-$3946 \citep{Uchi07}, 
where the year-scale time variability requires, most likely, 
the presence of a largely amplified magnetic field of 1~mG in compact filaments. 
The strength of magnetic field in  
the compact radio knots in Cas~A is estimated to 
be 1--3 mG \citep{Ato00},  just sufficient to produce year-scale variability through  
synchrotron cooling and shock acceleration of multi-TeV electrons. 
Indeed, here we report 
significant time variations in nonthermal X-ray emission from many 
compact regions. 
The distance to Cas~A is assumed to be 3.4 kpc \citep{Reed95} 
so the angular scale of $1\arcmin$ corresponds to 1~pc.

\section{Data}

We analyzed the X-ray data  obtained  with the ACIS-S3 chip 
on 2000 January 30 \citep{Hwa00,Gott01},
 2002 February 6 \citep{DR03}, 
 and 2004 February 8 \citep{Hwa04}, each with $\simeq 50$ ksec exposure. 
Table \ref{tbl:log} gives the log of the \emph{Chandra} observations. 
The observations  were made 
with ACIS-S3 in the graded mode, with which telemetry size can be reduced. 
The three observations have almost identical configuration such as 
the aim point and the roll angle, 
allowing us to directly compare the three images without taking account of 
the varying  point-spread functions (PSFs). 
The aim point is located $\sim 1.9\arcmin$ southeast 
of the central point source, the compact remnant left behind 
after the supernova explosion. 

The X-ray analysis was done using the CIAO software package (version 3.4).
We reprocessed the event data, 
applying standard data reduction with the latest calibration files.
We find that the sky coordinates of the central point source 
 in the three epochs are consistent with each other with an accuracy of  $\la 0.2\arcsec$.
The point source itself is estimated to be moving at an angular speed of 
$\simeq 0\farcs 02\ \rm yr^{-1}$ \citep{Tho01}, thus 
$0\farcs 08$ in the 4-yr span. 

\begin{deluxetable}{ccccccl}
\tabletypesize{\small}
\tablecaption{Log of \emph{Chandra} Observations\label{tbl:log}}
\tablewidth{0pt}
\tablehead{
\colhead{ID} & \colhead{OBS ID} & \colhead{Date} & 
\colhead{R.A.} & \colhead{Decl.} & \colhead{Roll} &
\colhead{Exposure} \\
\colhead{} & \colhead{} & \colhead{} & 
\colhead{(deg)} & \colhead{(deg)} & \colhead{(deg)} & \colhead{(ks)}  
}
\startdata
P1 & 114  & 2000 01 30 & 350.9176 & 58.7934 & 323.4 & 49.9 \\
P2 & 1952 & 2002 02 06 & 350.9172 & 58.7938 & 323.4 & 49.6\\
P3 & 5196 & 2004 02 08 & 350.9188 & 58.7949 & 325.5 & 49.5 
\enddata
\tablecomments{Sky coordinates are right ascension (R.A.) and declination (Decl.)
in J2000.0,  representing the pointing direction. 
The exposure time is the 
net integration time after standard filtering.}
\end{deluxetable}

\begin{figure}[htbp] 
\plotone{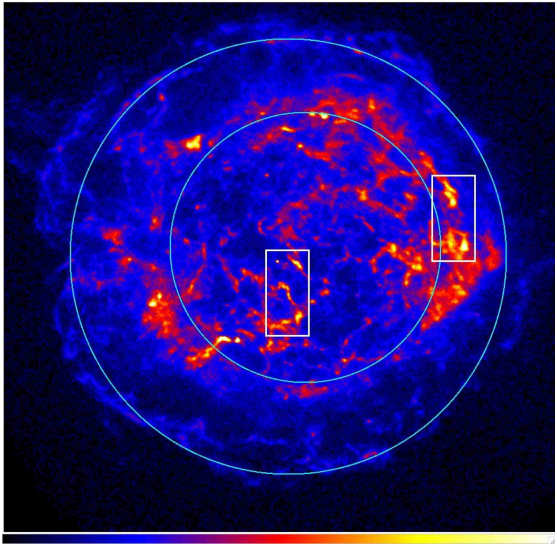}
\caption{\small 
\emph{Chandra} ACIS-S  image of Cas A 
in the  4--6 keV band averaged over the three epochs. 
North is up and east is to the left. 
The image is displayed with a square root scaling 
in a range of $0\mbox{--}3\ \rm counts\ pix^{-1}$.  
The outer circle with $153\arcsec$ radius denotes the forward shock positions, 
and the inner circle with $95\arcsec$ radius indicates the mean location 
of the reverse shock \citep{Gott01}.
\label{fig:whole} }
\end{figure}

Figure \ref{fig:whole} shows the 
\emph{Chandra}  image  (with $0\farcs2 \times 0\farcs2$ pixels) 
of Cas~A  in the 4--6 keV band averaged over the three epochs. 
This energy band corresponds to 
the continuum emission between Ca and Fe K lines, 
comprised of thermal bremsstrahlung and nonthermal component(s).
Assuming that the hard X-ray above 10 keV is dominated by 
nonthermal radiation processes, we estimate about $50\%$ of 4--6 keV photons
should be of nonthermal origin. 
Filamentary structures having 
high surface brightness in the continuum band 
are found only for radii less than $2\arcmin$ (referred as inner filaments),
implying their association with the reverse shock
as is shown to be the case by \citet{HV08}.

We are interested here in the nonthermal X-ray components.
We selected the two boxes shown in Fig.~\ref{fig:whole} 
based on the following criteria.
First, we identified regions sufficiently bright in the 4--6 keV band,  exceeding  
 $3\ \rm counts\ pix^{-1}$
at least in one epoch,  in order to obtain statistics enough for a variability search. 
Note that this condition is met only by the inner filaments, 
 associated with the reverse shock or its secondary shocks.
The second requirement is that 
the level of the continuum emission is high relative to the silicon line strength. 
Specifically, we calculated the ratio of 4--6 keV counts to 
1.7--2.2 keV counts at each pixel, and selected those  regions having 
the ratio exceeding 0.5. In this way, we found that the {\it inner filaments} 
satisfying these criteria are distributed inside the two boxes only. 
Since we do not aim at constructing a statistical sample in this work, 
we adopted these simple criteria, 
 accounting for neither position-dependent PSFs nor effective area. 

\section{Analysis}
\label{sec:ana}

\begin{figure*}[tbp]
\epsscale{0.5}
\plotone{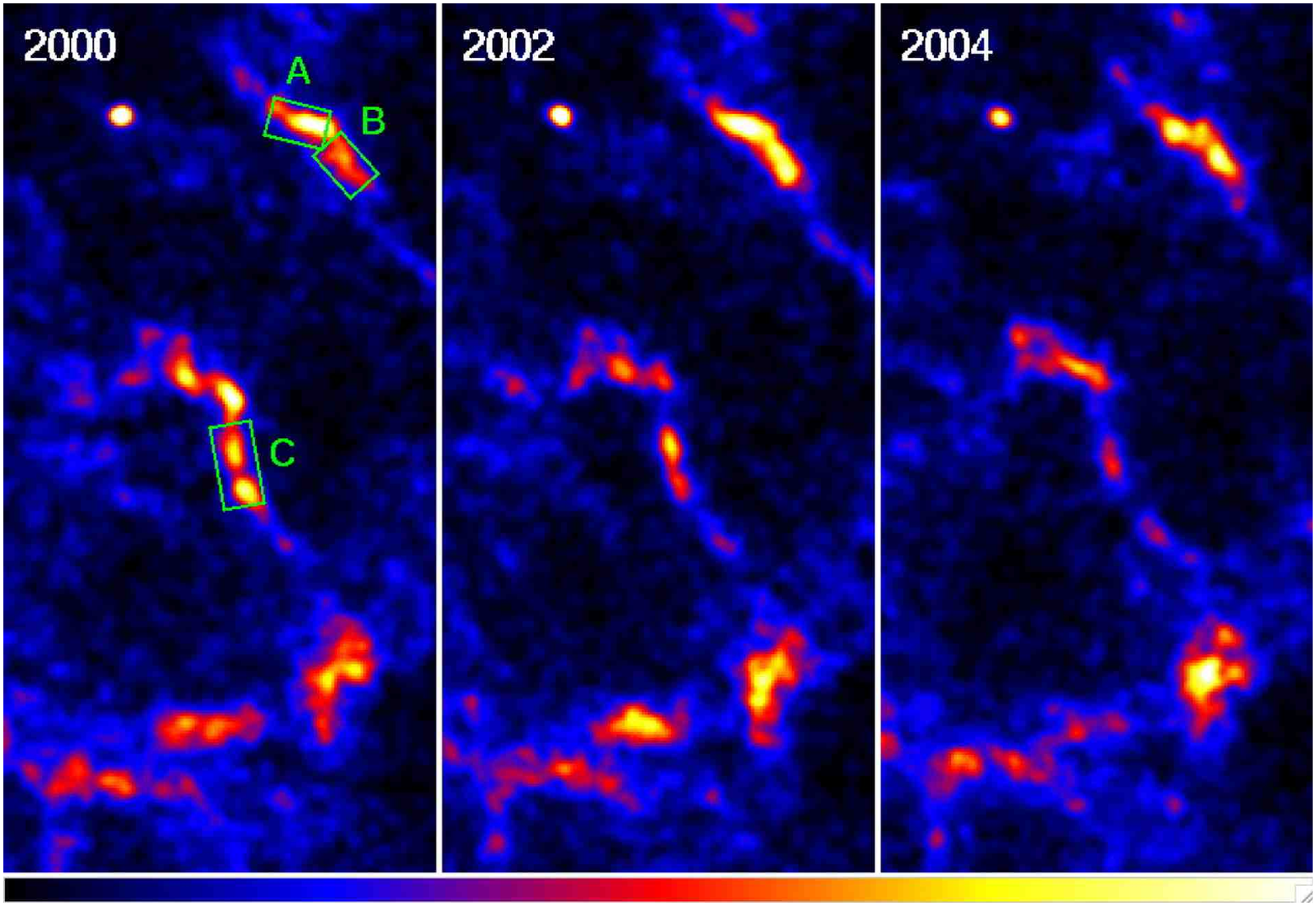}
\hspace{2mm}
\plotone{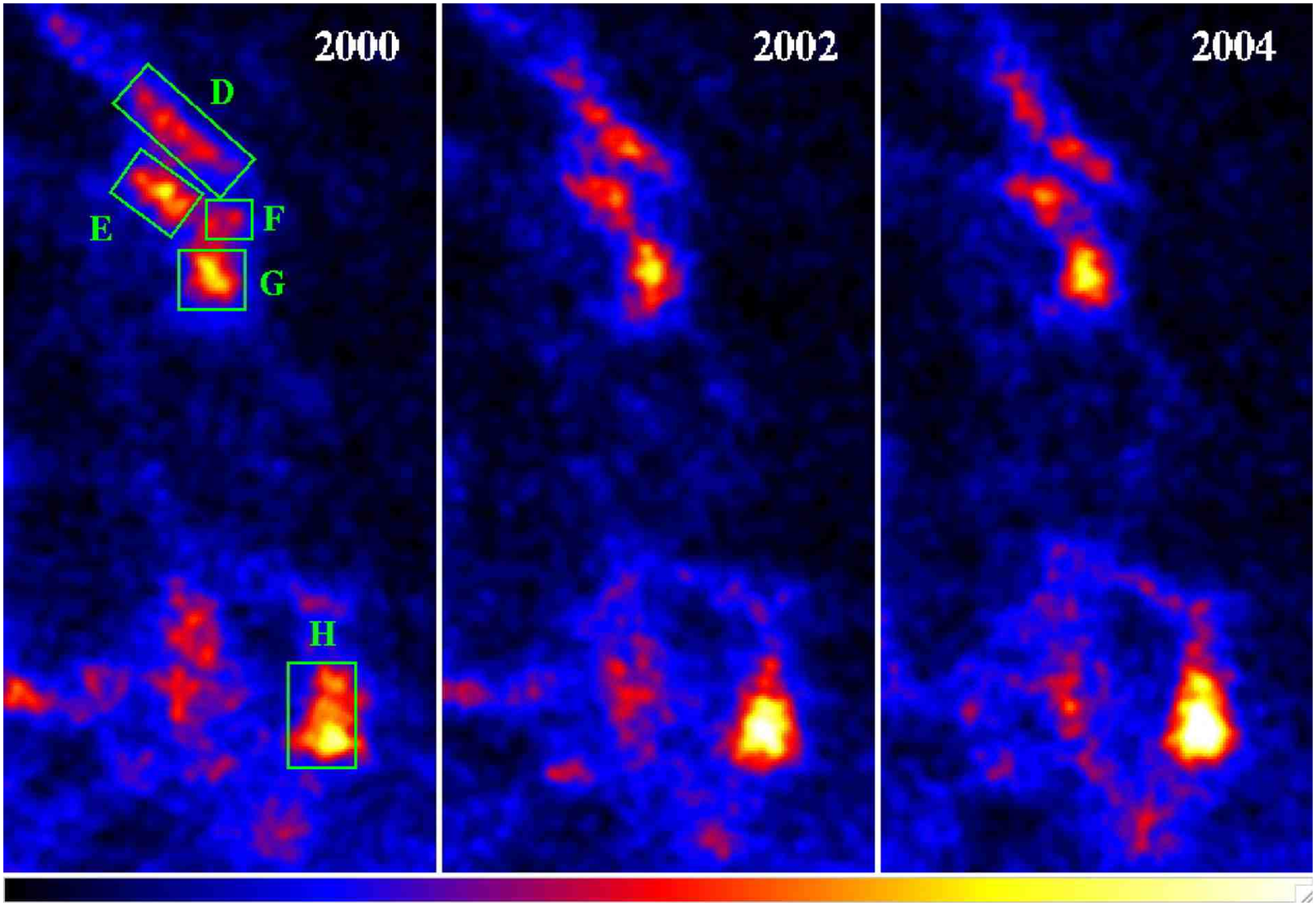}
\caption{\small  A sequence of  three-epoch 4--6 keV images of the 
two $0.5\arcmin \times 1\arcmin$ boxes in Fig.~\ref{fig:whole}. 
The images are shown in a linear scale in a range of 
$0\mbox{--}3\ \rm counts\ pix^{-1}$ for the left panel,  and 
$0\mbox{--}4\ \rm counts\ pix^{-1}$ for the right panel,  respectively. 
Pixels have dimensions of $0\farcs2 \times 0\farcs2$. 
Gaussian smoothing with a  kernel of $0\farcs8$ is applied. 
The central box (i.e., the left panel) is close to the aim point 
and therefore the PSF is sharp as is evident from the point source. 
The western box (the right panel) is away from the aim point, so that 
some of the spatial extent should be attributed to the broadening by the PSF. 
\label{fig:filaments} }
\end{figure*}

\begin{figure}[h]
\epsscale{1.2}
\plotone{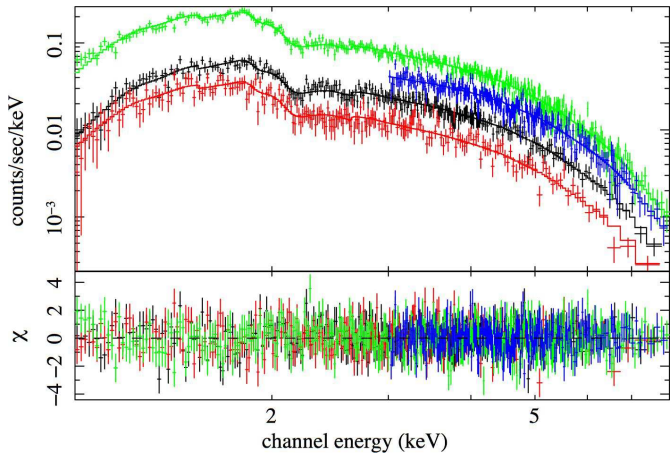}
\caption{\small  X-ray spectra of combined regions:
AB (black),  C (red), DEFG (green), and H (blue). 
The three epochs are averaged. 
The best-fit power-law function with interstellar attenuation is superimposed 
in each case. Nearby diffuse X-rays are subtracted from each source spectrum. 
\label{fig:xspec} }
\end{figure}

Figure \ref{fig:filaments} displays a sequence of 
three-epoch 4--6 keV images of the central box (left) and the western 
box (right), each with dimensions of  $0.5\arcmin \times 1\arcmin$.
Morphologically distinctive components that satisfy the criteria mentioned above 
are labeled as A, B, C, E, G, and H. Also we tagged an additional two features, 
D and F. 

The 4--6 keV images over a 4-yr time interval reveal a dozen  time-variable 
structures. 
Region C has decreased by a factor of 2 from 2000 to 2004. 
Region H underwent a flux brightening by 50\% over the 
time interval of 4 yr. 
 (By measuring the silicon line flux from the compact ejecta features, 
we estimated a systematic error in the flux measurements to be $< 9\%$.)
The variable filaments are spatially extended beyond \emph{Chanda}'s PSFs,  
typically $\sim 3\arcsec \mbox{--}5\arcsec$ in length and
$\sim 1\arcsec $ in width. 
Also, some filaments that we did not label,  such as those in the south of region C, 
exhibit time variability. 
Year-scale time variability appears to be the prevailing nature 
in the bright filaments or knots 
in the 4--6 keV band. 

Before investigating the epoch-dependence of the spectral parameters, 
we first constructed X-ray spectra averaged  over the three epochs. 
Regions A and B were combined into a single zone, and 
regions D, E, F, and G were merged. 
The background spectrum taken from a nearby, low brightness region 
was subtracted from each spectrum. 
Since the choice of a background region largely affects the 
low-energy part of the X-ray spectrum of region H, 
photons below 3 keV were discarded for this region. 
The four X-ray spectra (A$+$B, C, D$+$E$+$F$+$G, and H) 
in the 1--8 keV band are 
shown in Figure~\ref{fig:xspec}. They all appear to be line-less. 

To characterize the shape of the line-less X-ray spectrum, 
we adopted a simple power-law model attenuated  by interstellar absorption, 
which yielded statistically good fits.
We obtained the best-fit parameters in ranges of 
$\Gamma \simeq 2.0\mbox{--}2.4$ and 
 $N_{\rm H} = 1.7\mbox{--}2.0\times 10^{22}\ \rm cm^{-2}$ 
 as summarized in Table~\ref{tbl:parameter}.
The absorbing column density is broadly consistent with those for thermal components
\citep{Will02}.
We also performed spectral fitting using a thermal bremsstrahlung model. 
A statistically acceptable fit was obtained in each case. 
The best-fit electron temperature is in a range of 
4.0--6.6 keV, much higher than the electron temperatures,  1--2 keV,  derived for 
the X-ray knots showing thermal spectra \citep[e.g.,][]{LH03}.  
Moreover, by combining all the spectral data, 
the equivalent width of the iron K lines is constrained to be $<110$ eV. 
The absence of the emission lines 
 disfavors any bremsstrahlung interpretation of 
the featureless X-ray spectra, including bremsstrahlung by suprathermal 
electrons \citep{Lam01}.

\begin{deluxetable}{ccccl}
\tabletypesize{\small}
\tablecaption{Results of Spectral Fitting \label{tbl:parameter}}
\tablewidth{0pt}
\tablehead{
\colhead{ID} & \colhead{$N_{\rm H}$} & \colhead{$\Gamma$} & \colhead{$F_{2-10}$} 
& \colhead{$\chi^2_\nu ({\rm d.o.f.})$}\\
\colhead{} & \colhead{$10^{22}\ \rm cm^{-2}$} & \colhead{} & 
\colhead{$10^{-12} \rm erg\ cm^{-2}\ s^{-1}$} & \colhead{}
}
\startdata
AB & $2.00\pm 0.10$ & $2.31\pm 0.08$ & $1.59\pm 0.04$ & 0.85 (274)\\
C   & $1.80\pm 0.13$ & $2.37\pm 0.11$ & $0.77\pm 0.03$ & 1.08 (221)\\
DEFG & $1.69\pm 0.06$ & $2.24\pm 0.05$ & $5.21\pm 0.05$ & 0.98 (365) \\
H\tablenotemark{a} & 1.80 (fix) & $1.97\pm 0.10$ & $3.06\pm 0.07$ & 0.84 (206) 
\enddata
\tablecomments{Fitting X-ray spectrum by a power law with photoelectric absorption.
Absorbing column density $N_{\rm H}$, 
 photon index $\Gamma$, and unabsorbed 2--10 keV flux $F_{\rm 2-10}$ are 
shown  with 90\% errors.}
\tablenotetext{a}{X-ray energy below 3 keV is excluded from the fit and 
the absorbing column density is fixed at 
 $N_{\rm H} = 1.8\times 10^{22}\ \rm cm^{-2}$. }
\end{deluxetable}

\begin{figure}[htbp]
\epsscale{1.2}
\plotone{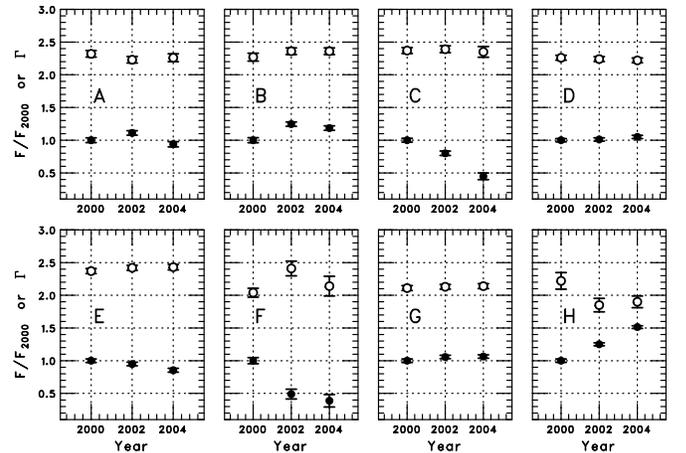}
\caption{\small  Time sequence of 
the unabsorbed 2--10 keV flux relative to that  in the year 2000 
for eight  filamentary regions (filled circles), 
as well as 
time sequence of photon index $\Gamma$ (open circles) are shown 
with $1\sigma$ errors.
\label{fig:evolution} }
\end{figure}

In order to track the time variations in terms of both intensity and spectral shape, 
the X-ray spectrum extracted from each labeled region at each epoch 
was fit by an absorbed power-law model. The absorbing column density 
was fixed at a relevant value in Table~\ref{tbl:parameter}.
Figure~\ref{fig:evolution} shows the time sequence of 
the unabsorbed X-ray flux normalized by that in the year 2000.
Four out of the eight regions exhibit large time variations ($>10\%\ \rm yr^{-1}$).
The time sequence of 
photon index $\Gamma$ for each region is also presented in Figure~\ref{fig:evolution}.
We found notable variations of $\Gamma$ only in two cases, regions F and H. 
Region H gives the hardest index of $\Gamma=1.85\pm 0.10$ in 2000, 
suggesting a direct link between flux increase and spectral hardening. 
The change in $\Gamma$ found in region F should be treated with care, 
because region F faded after 2000 and became almost invisible in 2002 and 2004. 
Photon index measured in 2002 and 2004  would not be physically connected 
with that in 2000. 

\section{Discussion}

We have found that a half ($4/8$) of the filamentary regions that 
were selected simply by their brightness and hardness, showed a
year-scale variability in X-ray flux. 
The X-ray spectra of the selected regions in the 1--8 keV range 
are featureless and well fit by an absorbed power law 
with a photon index $\Gamma \simeq 1.9\mbox{--}2.4$.
Given the absence of emission lines, we prefer to attribute 
the nonthermal X-ray emission in regions A--H to 
synchrotron radiation by very high-energy electrons. 
Synchrotron X-ray filaments or knots would originate in slightly overdense 
ejecta, say by a factor of $\sim 2$, into which a reverse shock (or transmitted shocks) 
with a typical shock speed of $\sim 1000 \rm \ km\ s^{-1}$ 
has been driven \citep[see][]{LH03}. 

Recently, \citet{PF07} have reported X-ray variability 
in compact thermal knots, resulting from the passage of reverse shock 
over the ejecta clumps. 
In the entire face of the remnant, they identified six time-varying structures; 
all but one region (their R4, corresponding to our region C) 
were modeled by a ``thermal" spectrum, specifically by a non-equilibrium-ionization 
plasma with an electron temperature of $\sim 1$ keV. 
On the other hand, 
we found a comparable number of variable ``nonthermal" features 
just inside the two small boxes, by making use of 4--6 keV maps which are 
sensitive to nonthermal variability. 
Morevoer, if we apply our criteria of significant variability, $>10\%\ \rm yr^{-1}$,  
to the \citet{PF07} sample, only two features remain. 
We conclude that 
variable nonthermal features are much more prevailing than the thermal ones. 

Cas~A is a very bright nonthermal 
radio source which indicates the presence in the remnant of 
relativistic electrons with a Lorentz factor $\gamma_{\rm e} \geq 100$.
The same electrons produce also high energy gamma-rays 
through relativistic bremsstrahlung. Since the gas density 
in Cas A is known quite well, the gamma-ray flux is 
robustly controlled by the strength of the magnetic field. Therefore, 
high energy gamma-ray observations, combined with 
radio data, can provide reliable measurements 
of the magnetic field \citep{Cowsik}. 
A simple homogeneous model based on the radio observations and the 
upper limits of the $\geq 100$~MeV flux reported by the 
EGRET collaboration \citep{Esp96} yields a lower 
limit on the magnetic field of 0.4 mG \citep{Ato00}.
In a more sophisticated two-zone model of \citet{Ato00}, 
the magnetic field in the compact radio knots, where
particle acceleration is supposed to be taking place, is constrained to be 
$B = 1\mbox{--}3\ \rm  mG$. These values would reasonably be applicable to 
the X-ray filaments, which have a similar size with the radio features, and 
are also presumed to be the sites of efficient particle 
acceleration by reverse (or secondary) shocks.

It is interesting to ask whether the X-ray flux changes are attributable 
solely to the changes in magnetic field strength. 
If so, it is expected that 
a substantial flux change in synchrotron X-rays accompanies a marked change in 
radio intensity. 
While the radio flux is proportional to $B^2$, the X-ray flux has weaker
dependence on $B$, because X-rays are emitted in a quasi-saturated regime. Therefore one would expect stronger variations in the radio than in the X-ray band.
According to \citet{AR95}, however, 
among about 70 bright radio compact knots exceeding $20\ \rm mJy\ beam^{-1}$, 
none showed a brightness increase larger than $+10\%\ \rm yr^{-1}$ 
in the archive Very Large Array data at 5 GHz. A typical rate of 
brightening is about $+3\%\ \rm yr^{-1}$, corresponding to an $e$-folding time 
of $\sim 30\ \rm yr$, which  is much longer than the X-ray variability timescale 
of $4\ \rm yr$. 
Therefore it is more likely that the observed time variations are caused mainly by 
a rapid increase or decrease in the density of X-ray-emitting electrons. 
(The increase in the radio-electron density above the   ``sea" level 
is expected to be less significant, 
 provided that fresh electrons have a harder energy spectrum
than old electrons.)
Indeed, as we argue below, the known constraint on $B$
provides the synchrotron cooling time 
being compatible with the observed time variability. 

We have reported a similar X-ray variability in SNR RX~J1713.7$-$3946,  and 
attributed it to fast synchrotron cooling and shock acceleration of electrons 
\citep{Uchi07}. Let us examine the observed variability timescale, using 
the equations presented in \citet{Uchi07}. 
Synchrotron cooling time is given by 
$t_{\rm syn}  \sim 1.5\, B_{\rm mG}^{-3/2} \varepsilon_{\rm keV}^{-1/2}\  \rm yr$, 
where $B_{\rm mG}$ is the magnetic field strength in units of mG and 
$\varepsilon_{\rm keV}$ is the mean synchrotron photon energy 
in units of keV.
For $B\sim 0.5\,\rm mG$, 
the synchrotron cooling timescale becomes comparable to the variability timescale, 
$t_{\rm syn} \sim t_{\rm var}  \sim 4$ yr.
Therefore, even a magnetic field of  0.4 mG 
is able to explain the observed variability. For $B \sim 2\ \rm mG$, 
if the injection of  relativistic electrons responsible for the synchrotron 
X-rays  has suddenly ceased in a filament, the X-ray flux can be reduced on 
a timescale of $t_{\rm syn} \sim 0.5\ \rm yr$.
If the electron injection rate (i.e., the number of particles 
picked up and injected into the acceleration process in unit time) decreases
noticeably in 4 yr, 
the number of X-ray-emitting electrons and consequently 
X-ray intensity can drop as observed. 
We note that plasma characteristics   may change on a timescale of  a few years 
given the arcsecond-scale size of the thermal ejecta filaments.

Rising flux is a signature of ongoing production (acceleration) of synchrotron-emitting electrons with energies of $\sim 10\ \rm TeV$. 
It implies that electron energy can increase to 
$\sim 10\ \rm TeV$ in 4 yr or less at the  shock.
Assuming that electrons are accelerated via diffusive shock acceleration, 
the acceleration time of the X-ray-emitting electrons is given by 
$t_{\rm acc}  \sim  9 \eta\, 
B_{\rm mG}^{-3/2}  
\varepsilon_{\rm keV}^{1/2} 
V_{1000}^{-2}\,  \rm yr$,
where $\eta \geq 1$ is a gyrofactor,  
and $V_{1000}$ is the velocity of the reverse shock relative to the unshocked ejecta 
ahead of it in units of $1000 \rm \ km\ s^{-1}$.
Therefore, 
reverse shocks 
with a typical speed of $\sim 1000 \rm \ km\ s^{-1}$ \citep{LH03} and 
a magnetic field of 2 mG can accelerate 
X-ray-emitting electrons 
within a timescale of $t_{\rm var}  \sim 4$ yr,  
provided that the acceleration proceeds  
close to the Bohm diffusion limit ($\eta \sim 1$). 

The presence of time variability together with 
the absence of the Fe-K lines strongly prefer 
a synchrotron origin of the hard X-ray emission in the selected filaments, 
providing new evidence of acceleration to multi-TeV energies at the reverse 
(and transmitted) shocks in ejecta. 
Though the regions we studied here contain only a portion of the reverse-shocked gas, 
the bulk of the hard X-rays above 10 keV 
is also likely of a synchrotron origin. If so, the steepness of the continuum spectrum 
above 10 keV, $\Gamma \sim 3$ \citep{Fava97}, indicates that 
the synchrotron spectrum rolls off  at around 1 keV, 
being consistent with the theory of diffusive shock 
acceleration \citep[see][]{Uchi07}. 
The maximum energy of the electrons is thus constrained to be  $\sim 10\ \rm TeV$.
 Therefore the TeV gamma-rays measured  by 
HEGRA \citep{Aha01} and 
 MAGIC \citep{Magic} groups are difficult to be explained by electron 
 bremsstrahlung in the reverse-shocked ejecta, 
 and are likely to be of a hadronic origin.

The interpretation of the X-ray variability has to be critically assessed 
through future theoretical and observational works. 
Regardless of what 
is behind 
the nonthermal X-ray variability, it offers us a new diagnostic tool to 
understand particle acceleration in supernova shocks. 

\acknowledgments

We wish to thank Herman Lee for his useful comments on 
the original manuscript. 







\begin{thebibliography}{}

\bibitem[Aharonian et al.(2001)]{Aha01} Aharonian, F., et al.\ 
(HEGRA collaboration) 2001, \aap, 370, 112 
\bibitem[Albert et al.(2007)]{Magic} Albert, J. et al.\
(MAGIC collaboration) 2007, \aap, 474, 937
\bibitem[Allen et al.(1997)]{Allen97} Allen, G.~E., et al.\ 1997, \apjl, 487, L97
\bibitem[Anderson \& Rudnick(1995)]{AR95} Anderson, M.~C., 
\& Rudnick, L.\ 1995, \apj, 441, 307 
\bibitem[Atoyan et al.(2000)]{Ato00} Atoyan, A.~M., 
Aharonian, F.~A., Tuffs, R.~J., V\"olk, H.~J.\ 2000, \aap, 355, 211 
\bibitem[Baars et al.(1977)]{Baars77} Baars, J.~W.~M., Genzel, 
R., Pauliny-Toth, I.~I.~K., \& Witzel, A.\ 1977, \aap, 61, 99
\bibitem[Bleeker et al.(2001)]{Blee01} Bleeker, J.~A.~M., 
Willingale, R., van der Heyden, K., Dennerl, K., Kaastra, J.~S., 
Aschenbach, B., \& Vink, J.\ 2001, \aap, 365, L225 
\bibitem[Cowsik \& Sarkar (1980)]{Cowsik} Cowsik, R., \& Sarkar, S.
1980, MNRAS, 191, 855 
\bibitem[DeLaney \& Rudnick(2003)]{DR03} DeLaney, T., \& 
Rudnick, L.\ 2003, \apj, 589, 818 
\bibitem[Esposito et al.(1996)]{Esp96} Esposito, J.~A., 
Hunter, S.~D., Kanbach, G., \& Sreekumar, P.\ 1996, \apj, 461, 820 
\bibitem[Favata et al.(1997)]{Fava97} Favata, F., et al.\ 1997, \aap, 324, L49 
\bibitem[Gotthelf et al.(2001)]{Gott01} Gotthelf, E.~V., 
Koralesky, B., Rudnick, L., Jones, T.~W., Hwang, U., \& Petre, R.\ 2001, 
\apjl, 552, L39 
\bibitem[Helder \& Vink(2008)]{HV08} Helder, E. A., \& Vink, J.\ 2008, \apj, submitted 
\bibitem[Holt et al.(1994)]{Holt94} Holt, S.~S., Gotthelf, 
E.~V., Tsunemi, H., \& Negoro, H.\ 1994, \pasj, 46, L151 
\bibitem[Hughes et al.(2000)]{Hugh00} Hughes, J.~P., Rakowski, 
C.~E., Burrows, D.~N., \& Slane, P.~O.\ 2000, \apjl, 528, L109 
\bibitem[Hwang et al.(2000)]{Hwa00} Hwang, U., Holt, S.~S., 
\& Petre, R.\ 2000, \apjl, 537, L119 
\bibitem[Hwang et al.(2004)]{Hwa04} Hwang, U., et al.\ 2004, 
\apjl, 615, L117 
\bibitem[Laming(2001)]{Lam01} Laming, J.~M.\ 2001, \apj, 546, 1149
\bibitem[Laming \& Hwang(2003)]{LH03} Laming, J.~M., \& Hwang, U.\ 2003, \apj, 597, 347
\bibitem[Patnaude \& Fesen(2007)]{PF07} Patnaude, D. J., \& Fesen, R. A.\
2007, \aj, 133, 147 
\bibitem[Reed et al.(1995)]{Reed95} Reed, J. E., Hester, J. J., Fabian, A. C., 
\& Winkler, P. F.\ 1995, \apj, 440, 706 
\bibitem[Thorstensen et al.(2001)]{Tho01} Thorstensen, J.~R., 
Fesen, R.~A., \& van den Bergh, S.\ 2001, \aj, 122, 297 
\bibitem[Uchiyama et al.(2007)]{Uchi07} Uchiyama, Y., 
Aharonian, F.~A., Tanaka, T.,  Takahashi, T., \& Maeda, Y.\ 2007, \nat, 449, 576 
\bibitem[Vink \& Laming(2003)]{VL03} Vink, J., \& Laming, J.~M.\ 2003, \apj, 584, 758
\bibitem[Willingale et al.(2002)]{Will02} Willingale, R., 
Bleeker, J.~A.~M., van der Heyden, K.~J., Kaastra, J.~S., \& Vink, J.\ 
2002, \aap, 381, 1039 
\end{thebibliography}
\end{document}